\documentclass{IEEEtran4PSCC}
\ifCLASSINFOpdf
   \usepackage[pdftex]{graphicx}
\else
   \usepackage[dvips]{graphicx}
\fi
\usepackage[cmex10]{amsmath}
\usepackage{algorithmic}

\usepackage{tabularx, booktabs}
\usepackage[ruled,vlined]{algorithm2e}
\usepackage{amsfonts}
\usepackage{multirow}
\usepackage[table,xcdraw]{xcolor}

\usepackage{tikz}

\newcommand{\Acal}{{\cal A}}

\newcommand{\Kcal}{{\cal K}}

\newcommand{\Mcal}{{\cal M}}

\newcommand{\Pcal}{{\cal P}}

\newcommand{\Scal}{{\cal S}}

\newcommand{\argmin}{\mathop{\rm argmin}}
\newcommand{\argmax}{\mathop{\rm argmax}}

\newtheorem{remark}{Remark}

\makeatletter
\let\old@ps@headings\ps@headings
\let\old@ps@IEEEtitlepagestyle\ps@IEEEtitlepagestyle
\def\psccfooter#1{%
    \def\ps@headings{%
        \old@ps@headings%
        \def\@oddfoot{\strut\hfill#1\hfill\strut}%
        \def\@evenfoot{\strut\hfill#1\hfill\strut}%
    }%
    \def\ps@IEEEtitlepagestyle{%
        \old@ps@IEEEtitlepagestyle%
        \def\@oddfoot{\strut\hfill#1\hfill\strut}%
        \def\@evenfoot{\strut\hfill#1\hfill\strut}%
    }%
    \ps@headings%
}
\makeatother

\usepackage{cite}

\makeatletter
\renewcommand\Huge{\@setsize\Huge{25pt}\xxvpt{22.25pt}}
\renewcommand{\baselinestretch}{0.98} 
\makeatother

\begin{document}
%
\title{A Reinforcement Learning Approach to Parameter Selection for Distributed Optimal Power Flow}

\author{
\IEEEauthorblockN{Sihan Zeng, Daniel K. Molzahn}
\IEEEauthorblockA{School of Electrical and Computer Engineering\\
Georgia Institute of Technology\\
Atlanta, Georgia USA\\
\{szeng30, molzahn\}@gatech.edu}
\and
\IEEEauthorblockN{Alyssa Kody$^*$, Youngdae Kim$^\dagger$, Kibaek Kim$^\dagger$}
\IEEEauthorblockA{Energy Systems$^*$, Mathematics and Computer Science$^\dagger$ \\
Argonne National Laboratory\\
Lemont, Illinois USA\\
\{akody, youngdae, kimk\}@anl.gov}
}


\maketitle

\begin{abstract}
With the increasing penetration
of distributed energy resources, distributed optimization algorithms have attracted significant attention for power systems applications due to their potential for superior scalability, privacy, and robustness to a single point-of-failure. The Alternating Direction Method of Multipliers (ADMM) is a popular distributed optimization algorithm; however, its convergence performance is highly dependent on the selection of penalty parameters, which are usually chosen heuristically.  
In this work, we use reinforcement learning (RL) to develop an adaptive penalty parameter selection policy for the AC optimal power flow (ACOPF) problem solved via ADMM with the goal of minimizing the number of iterations until convergence. 
We train our RL policy using deep Q-learning, and show that this policy can result in significantly accelerated convergence (up to a 59\% reduction in the number of iterations compared to existing, curvature-informed penalty parameter selection methods).
Furthermore, we show that our RL policy demonstrates promise for generalizability, performing well under unseen loading schemes as well as under unseen losses of lines and generators (up to a 50\% reduction in iterations).
This work thus provides a proof-of-concept for using RL for parameter selection in ADMM for power systems applications.
\end{abstract}



\section{Introduction}

The rapid growth in distributed energy resources (DER) such as solar PV, batteries, and plug-in vehicles necessitates new computational methods for cooperatively controlling these devices to maximize the efficiency and reliability of power systems. Traditionally, set-points for controllable devices are determined using centralized methods, meaning that all data are congregated in a central location (often an independent system operator), which solves a large-scale optimization problem. However, centralized methods may be unable to computationally manage the increase in problem size and complexity resulting from adding millions of DERs. Furthermore, centralized computing raises other issues such as data privacy \cite{kroposki2020autonomous,ryu2021privacy} and is also vulnerable via a single point of failure or attack. Consequently, there has been much interest from the power systems community in distributed computation methods, where large problems are partitioned into smaller problems that can be solved in parallel \cite{molzahn2017survey}. Distributed optimization can be used to either (i) physically spread computation across an electric network such that devices locally solve a small optimization problem and exchange solutions directly with neighboring devices until converging to the overall solution~\cite{wang2017distributed}, or (ii) partition large problems in the context of high-performance computing (HPC)~\cite{ExaTron}.

Despite the promise of distributed methods for power systems applications, they have not yet been widely adopted in industry. 
A review by Wang et al.~\cite{wang2017distributed} finds that one reason for this lack of adoption is that distributed optimization algorithms ``may require many iterations and in turn increase computational burden beyond the limit of practical interest for power industry.''
For example, commonly used distributed methods in power systems such as Alternating Direction Method of Multipliers (ADMM), Auxiliary Problem Principle (APP), and Analytical Target Cascading (ATC) may require hundreds or thousands of iterations to converge to a sufficiently high accuracy and only have convergence guarantees for a limited class of problems \cite{molzahn2017survey, wang2017distributed, kargarian2016toward}. 

While the worst-case computational performance of optimization algorithms is characterized by complexity theory, in practice, user-selected algorithmic parameters can significantly reduce typical solution times.
For example, it is widely known that the convergence performance of ADMM is sensitive to the choice of penalty parameters, which are heuristically defined~\cite{boyd2011distributed}. Furthermore, poor parameter selection can lead to solution divergence. 
In \cite{mhanna2018component}, Mhanna et al. demonstrate that nonconvex and nonlinear alternating current optimal power flow (ACOPF) problems~\cite{bienstock2015nphard} solved via ADMM have widely varying convergence results based on the selection of penalty parameters, including divergence.
Recent theoretical advancements \cite{sun2021two} enable ADMM to guarantee convergence for the ACOPF problem; however, convergence performance still depends on parameter settings. 

To speed up convergence and reduce the effort of penalty parameter tuning in ADMM, adaptive penalty parameter algorithms have been studied in order to update penalty parameters during the optimization using feedback from the previous iteration.
Examples include residual balancing \cite{he2000alternating}, which increases or decreases penalty parameters based on the relative magnitudes of the primal and dual residuals, and methods that use estimates of the local curvature of the dual function to inform updates \cite{xu2017adaptive}.
Mhanna et al. in \cite{Mhanna19} demonstrate significantly improved convergence performance for the ACOPF problem using adaptive penalty parameter algorithms over vanilla ADMM with static penalty parameters.
However, the techniques in \cite{Mhanna19} still rely on tuned parameters within the adaptive algorithm, and also require  additional logic steps and the computation and storage of gradient information.

Ultimately, these existing adaptive penalty parameter algorithms rely on heuristics, presenting an opportunity for their replacement with machine learning techniques that may have superior performance.
In this work, we develop a reinforcement learning (RL) \cite{chen2021reinforcement} method to train a policy for selecting penalty parameters to accelerate the convergence of an ADMM algorithm for solving ACOPF problems. The ADMM parameter selection task has a sequential decision making structure, as penalty parameters are updated based on feedback from past iterations. RL, as a convenient framework for sequential decision making problems, is a natural fit for this task.


Machine learning techniques have been used to design optimization methods (e.g., \cite{chen2021learning,andrychowicz2016learning}). 
There are fewer works that develop embedded-ML methods specifically for distributed optimization algorithms. In \cite{biagioni2020learning}, a recurrent neural network is trained to predict the converged values of variables in ADMM subproblems for DC-OPF. In \cite{graf2019distributed}, the authors replace ADMM subproblems with an RL policy that predicts solutions. In \cite{xie2019differentiable}, the authors learn to solve ADMM subproblems by recasting them as deep neural networks. 
Recent contemporaneous work \cite{ichnowski2021accelerating} trains an RL policy to tune parameters to accelerate ADMM convergence using policy gradient methods; however, they focus on convex QP problems with convergence guarantees and do not specifically consider power systems problems.
Moreover, RL methods have shown promise in other power systems applications (e.g.,~\cite{li2018alphago,duchesne2020recent}).

In this work, we investigate the use of RL in the important task of learning ADMM penalty parameters. Transforming the penalty parameter selection problem into an RL problem, this work has three main contributions:
\begin{itemize}
    \item Formulation of parameter selection for distributed ACOPF as a RL problem.
    \item Development of a novel deep Q-learning policy scheme for ADMM penalty parameter selection.
    \item Demonstration of trained policies on test networks with unseen loads and unseen line and generator contingencies.
\end{itemize}

The rest of the paper is organized as follows. In Section \ref{sec:formulation}, we introduce the ACOPF problem and its component-based ADMM formulation. We discuss the importance of the penalty parameter $\rho$ for the convergence of the ADMM algorithm. In Section \ref{sec:background}, we briefly highlight the connection between the penalty parameter selection problem and RL and provide an overview of RL and deep Q-learning. In Section \ref{sec:method}, we dive into our RL algorithm design, including the choice of the state space, action space, and reward function. We present numerical experiments in Section \ref{sec:experiments}, and conclude with future directions in Section \ref{sec:conclusion}.

\section{Component-based Decomposition of ACOPF}\label{sec:formulation}

We consider a component-based decomposition of ACOPF~\cite{mhanna2018component,Mhanna19} that can be efficiently solved by ADMM, where each component in the network (i.e., buses, lines, generators) form their own subproblems.
Although region-based ADMM decompositions \cite{sun2021two, erseghe2014distributed} are also popular for power systems applications and result in fewer subproblems, the advantage of the component-based formulation is that each subproblem is small and can be solved efficiently, lending itself well to HPC implementations~\cite{ExaTron}.
Furthermore, component-based decompositions do not require making partitioning decisions, which can impact performance.


\subsection{ACOPF Formulation}
\label{subsec:acopf}

We present the ACOPF problem formulation below in~\eqref{eq:acopf}. This problem seeks the least costly operating points of the generators within their lower and upper limits while obeying physical laws. These physical laws are represented by power flow equations~\eqref{eq:pg}--\eqref{eq:qg} and~\eqref{eq:pij}--\eqref{eq:qji}.
%
\begin{subequations}
\begin{align}
    &\underset{p_{g_i},q_{g_i},w_i,\theta_i,w^R_{ij},w^I_{ij}}{\text{minimize}} \; \sum_{i \in \mathcal{B}}\sum_{g_i \in \mathcal{G}_i} f_{g_i}(p_{g_i}) \hspace*{-3em}\\
    &\text{subject to} \notag\\
    &\sum_{g_i \in \mathcal{G}_i} p_{g_i} - p_{d_i} = g_i^Sw_i + \sum_{j \in \mathcal{B}_i}p_{ij}, \hspace*{-3em} & \forall i \in \mathcal{B} \label{eq:pg}\\
    &\sum_{g_i \in \mathcal{G}_i} q_{g_i} - q_{d_i} = -b_i^Sw_i + \sum_{j \in \mathcal{B}_i}q_{ij}, \hspace*{-3em} & \forall i \in \mathcal{B} \label{eq:qg}\\
    &\sqrt{p_{ij}^2 + q_{ij}^2} \le \bar{r}_{ij}, & \forall (i,j) \in \mathcal{L} \label{eq:ij-limit}\\
    &\sqrt{p_{ji}^2 + q_{ji}^2} \le \bar{r}_{ij}, & \forall (i,j) \in \mathcal{L} \label{eq:ji-limit}\\
    & \underline{p}_{g_i} \le p_{g_i} \le \overline{p}_{g_i}, & \forall g_i \in \mathcal{G}_i, \forall i \in \mathcal{B}\\
    & \underline{q}_{g_i} \le q_{g_i} \le \overline{q}_{g_i}, & \forall g_i \in \mathcal{G}_i, \forall i \in \mathcal{B}\\
    & -2\pi \le \theta_i \le 2\pi, & \forall i \in \mathcal{B}\\
    & p_{ij} = g_{ii}w_i + g_{ij}w^R_{ij} + b_{ij}w^I_{ij}, & \forall (i,j) \in \mathcal{L} \label{eq:pij}\\
    & q_{ij} = -b_{ii}w_i - b_{ij}w^R_{ij} + g_{ij}w^I_{ij}, & \forall (i,j) \in \mathcal{L}\label{eq:qij}\\
    & p_{ji} = g_{jj}w_j + g_{ji}w^R_{ij} - b_{ji}w^I_{ij}, & \forall (i,j) \in \mathcal{L} \label{eq:pji}\\
    & q_{ji} = -b_{jj}w_j - b_{ji}w^R_{ij} - g_{ji}w^I_{ij}, & \forall (i,j) \in \mathcal{L} \label{eq:qji}\\
    & (w^R_{ij})^2 + (w^I_{ij})^2 = w_iw_j, & \forall (i,j) \in \mathcal{L} \label{eq:wij-magnitude}\\
    & \theta_i - \theta_j = \arctan(w^I_{ij}/w^R_{ij}), & \forall (i,j) \in \mathcal{L} \label{eq:wij-angle}
\end{align}
\label{eq:acopf}
\end{subequations}

In this optimization problem, we use $\mathcal{B},\mathcal{B}_i,\mathcal{G}_i$, and $\mathcal{L}$ to denote
the set of buses, the set of buses connected to bus $i$, the set of generators at bus $i$, and the set of lines, respectively. The decision variables include $p_{g_i}$ and $q_{g_i}$, which are the real and reactive power outputs of generator $g_i$ at bus $i$, $w_i$ and $\theta_i$, which are the squared voltage magnitude $(=v_i^2)$ and angle at bus $i$, and $w^R_{ij}$ and $w^I_{ij}$, which are defined to be $v_iv_j\cos\theta_{ij}$ and $v_iv_j\sin\theta_{ij}$, respectively, with $v_i$ being the voltage magnitude at bus $i$ and $\theta_{ij}:=\theta_i-\theta_j$. This choice of problem formulation is more naturally suited to the component-based decomposition that we adopt from~\cite{ExaTron} in this work. $f_{g_i}(\cdot)$ is a quadratic function of the real power output that encodes the power generation cost. The other quantities in \eqref{eq:pg}--\eqref{eq:wij-magnitude} are parameters that depend on the structure and physical properties of the power network (see \cite{Mhanna19} for more details).




\subsection{ADMM Formulation for ACOPF}
\label{subsec:admm-formulation}

Consider the following optimization, which is the general problem form for ADMM:
\begin{equation}
    \begin{aligned}
    & \underset{x\in\mathbb{R}^{n_1},\bar{x}\in\mathbb{R}^{n_2}}{\text{minimize}} && f(x)+g(\bar{x})\\
    & \text{subject to} && Ax+B\bar{x}=c,
    \end{aligned}
    \label{eq:admm}
\end{equation}
where $A\in\mathbb{R}^{n_3\times n_1}$, $B\in\mathbb{R}^{n_3\times n_2}$, and $c\in\mathbb{R}^{n_3}$, and where $f: \mathbb{R}^{n_1} \rightarrow \mathbb{R}$ and $g: \mathbb{R}^{n_2} \rightarrow \mathbb{R}$ are closed functions.
Let $y \in \mathbb{R}^{n_3}$ be the vector of Lagrange multipliers used to enforce the linear equality constraint in \eqref{eq:admm}.
Then, we form the augmented Lagrangian as
$L_\rho(x,\bar{x},y)=f(x)+g(\bar{x})+y^T(Ax+B\bar{x}-c)+(Ax+B\bar{x}-c)^{\top}\Omega(Ax+B\bar{x}-c)$, where the matrix $\Omega\in\mathbb{R}^{n_3\times n_3}$ is a diagonal matrix with the $i$-th diagonal entry defined as $\Omega_{ii}=\rho_i/2$. 
We define $\rho_i$ as the $i$-th \textit{penalty parameter}.

Let $k \in \mathbb{N}$ be the ADMM iteration counter, where iterates are marked via square brackets in superscript. 
Each iteration, we first update variable $x$ according to \eqref{eq:admm_update_1}.
Then, using this updated value of $x$, variable $\bar{x}$ is updated according to \eqref{eq:admm_update_2}.
Finally, the Lagrange multipliers are updated via \eqref{eq:admm_update_3}.
\begin{subequations} \label{eq:admm_update}
\begin{align}
    &x^{[k+1]}=\argmin_{x}L_{\rho}(x,\bar{x}^{[k]},y^{[k]}) \label{eq:admm_update_1}\\
    &\bar{x}^{[k+1]}=\argmin_{\bar{x}}L_{\rho}(x^{[k+1]},\bar{x},y^{[k]})\label{eq:admm_update_2}\\
    &y^{[k+1]}=\argmin_{y}L_{\rho}(x^{[k+1]},\bar{x}^{[k+1]},y) \label{eq:admm_update_3}
\end{align}
\end{subequations}
This iterative process continues until the 2-norms of the primal and dual residuals, which represent the feasibility of the primal and dual problems, have met their convergence thresholds, $\epsilon_{p}>0$ and $\epsilon_{d}>0$, respectively:
\begin{align}
    \left\|r^{[k]}_{p}\right\|_{2}\leq\epsilon_{p} \quad\text{and}\quad\left\|r^{[k]}_{d}\right\|_{2}\leq\epsilon_{d}, \label{eq:convergence}
\end{align}
where $r^{[k]}_{p}$ and $r^{[k]}_{d}$ are the  primal and dual residuals:
\begin{align}
    r^{[k]}_{p} &= Ax^{[k]} + B\bar{x}^{[k]} - c \label{eq:prim_res}\\
    r^{[k]}_{d} &= 2 \Omega A^TB \left( \bar{x}^{[k]} - \bar{x}^{[k-1]}\right). \label{eq:dual_res}
\end{align}

Reference~\cite{Mhanna19} proposes a method to decompose the ACOPF problem~\eqref{eq:acopf}, based on the observation that components can be decoupled by duplicating variables connecting them.
Generators and buses are coupled through the $p_{g_i}$ and $q_{g_i}$ variables, and branches and buses are coupled through the $p_{ij},q_{ij},p_{ji},q_{ji},w_i,\theta_i,w_j,$ and $\theta_j$ variables for a given branch $(i,j)$.
By duplicating these variables and enforcing a consensus through coupling constraints, we can reformulate the problem as the composition of small sub-problems, which can be written in the form of~\eqref{eq:admm} with proper choices of $A, B$ and $c=0$.
Applying ADMM to the reformulation permits massively parallel computations that can be accelerated using GPUs. In this work, we use the GPU-based solver developed in \cite{ExaTron}, which has achieved the state-of-the-art performance in terms of computation speed for solving ADMM problems.

Note that the $i$-th coupling constraint in the ADMM formulation is associated with penalty parameter $\rho_i$.
In \cite{mhanna2018component}, improved convergence performance is observed for ACOPF when $\rho_i$ values are assigned based on the type of coupling constraint they are penalizing.
Therefore, we categorize the coupling constraints into two different types: constraints that correspond to the real ($p$) and reactive ($q$) power flows, and constraints that correspond to voltage magnitudes ($v$) and angles ($\theta$). We use $n_{pq}$ and $n_{v\theta}$ to denote the number of the two types of constraints, respectively.
We use $\rho_{pq}\in\mathbb{R}^{n_{pq}}$ for the penalty parameters for the $p$ or $q$ coupling constraints and $\rho_{v\theta}\in\mathbb{R}^{n_{v\theta}}$ for the penalty parameters for the $v$ or $\theta$ coupling constraints.

\section{Reinforcement Learning Overview}\label{sec:background}

From the perspective of accelerating convergence, we seek the optimal parameter $\rho$ throughout the ADMM iterations to encourage the primal and dual residuals to reach the convergence thresholds in as few iterations as possible. The choice of $\rho$ in the $k$-th ADMM iteration is based on the current iterates $x^{[k]},\bar{x}^{[k]},y^{[k]}$, and in turns affects $x^{[k+1]},\bar{x}^{[k+1]},y^{[k+1]}$, the iterates of the next iteration. This naturally makes the problem a sequential decision making problem, which motivates us to approach it using RL. 
In this section, we provide an overview of RL modeled as a Markov Decision Process (MDP) and discuss Q-learning, an effective class of RL algorithms that we will use in this work.

\subsection{Reinforcement Learning \& Markov Decision Process}

Reinforcement learning is a framework for sequential decision making that involves an agent interacting with an environment. The agent observes the state and reward information from the environment and selects an action in response. The action makes the environment transition from the current state to the next state and reveal the next reward. The goal of the agent is to choose the optimal actions to maximize the discounted cumulative reward it receives from the environment.

Mathematically, we consider the Markov Decision Process (MDP), characterized by the 5-tuple $\Mcal=(\Scal,\Acal,\Pcal,R,\gamma)$.
$\Scal$ and $\Acal$ denote the state and action space. $\Pcal:\Scal\times\Acal\rightarrow\Delta_{\Scal}$ (with $\Delta_{\Kcal}$ denoting the probability simplex over a set $\Kcal$) is the transition probability kernel that specifies the distribution of the next state given the current state and action. $R:\Scal\times\Acal\rightarrow\mathbb{R}$ is the reward function. $\gamma\in(0,1)$ is the discount factor that discounts rewards received in the future. 
Due to the Markovian nature of the environment, selecting the optimal sequence of actions can be equivalently expressed as finding a policy $\pi:\Scal\rightarrow\Delta_{\Acal}$. The policy is a mapping from the state space to the probability simplex over the action space, and we use $\pi(a\mid s)$ to represent the probability of choosing action $a$ in state $s$.
The RL agent seeks to maximize the discounted cumulative reward by solving the optimization problem
\begin{align}
    \max_{\pi}~ &\mathbb{E}\left[\sum_{k=0}^{\infty}\gamma^{k}R(s^{[k]},a^{[k]})\right]\label{eq:RL_objective}\\
    \text{s.t. }& a^{[k]}\hspace{-2pt}\sim\hspace{-2pt}\pi(\cdot\mid s^{[k]}),\,\,s^{[k+1]}\hspace{-2pt}\sim\hspace{-2pt}\Pcal(\cdot\mid s^{[k]},a^{[k]}),\,\forall k=0,1,\ldots\notag
\end{align}
where $x\sim d$ denotes drawing a sample $x$ uniformly from the distribution $d$.

Two main classes of methods to solve the RL problem \eqref{eq:RL_objective} are the policy gradient algorithm and Q-learning. The method used in this work is a variant of Q-learning, which we briefly review in the following subsection. 

\subsection{Q-Learning}

In RL, the ``value" of a state-action pair under a policy $\pi$ is measured by the discounted cumulative reward obtained by applying action $a$ in state $s$ and then following the policy~$\pi$:
\begin{align*}
    Q_{\pi}(s,a)=\mathbb{E}_{\pi}\left[\sum_{k=1}^{\infty}\gamma^{k}R(s^{[k]},a^{[k]})\mid s^{[0]}=s, a^{[0]}=a\right].
\end{align*}
This is commonly known as the Q function under policy $\pi$. Under mild assumptions on the reward function, there always exists a deterministic optimal policy $\pi^*$ \cite{Puterman_book_1994}, which has a Q function obeying the Bellman equation for all $s\in\Scal$, $a\in\Acal$:
\[
Q_{\pi^*}(s,a)=R(s,a)+\gamma\mathbb{E}_{s'\sim\Pcal(\cdot\mid s,a)}[\max_{a'\in\Acal}Q_{\pi^*}(s',a')].
\]
On the other hand, $\pi^*$ can be determined from its Q function. Defining $a^*(s)=\argmax_{a\in\Acal}Q_{\pi^*}(s,a)$, we have
\begin{align*}
    \pi^*(a \mid s)=\begin{cases} 1, & \text { if } a=a^*(s), \\ 0, & \text { otherwise. }\end{cases}
\end{align*}
In other words, the optimal policy $\pi^*$ is to always take the action with the largest Q value. This suggests that to learn $\pi^*$, we can equivalently learn its Q function through stochastic approximation \cite{robbins1951stochastic}, where we maintain a table $Q^{[k]}\in\mathbb{R}^{|\Scal|\times|\Acal|}$ to track $Q_{\pi^*}$ and update it iteratively as
\begin{align*}
    Q^{[k+1]}\hspace{-2pt}\left(s^{[k]}, a^{[k]}\right)\hspace{-2pt}&=Q^{[k]}\left(s^{[k]}, a^{[k]}\right)+\alpha^{[k]}\Big(R\left(s^{[k]},a^{[k]}\right)\\
    &\hspace{-10pt}+\gamma \max_{a \in \Acal} Q^{[k]}\hspace{-2pt}\left(s^{[k+1]}, a\right)\hspace{-2pt}-Q^{[k]}\left(s^{[k]}, a^{[k]}\right)\Big),
\end{align*}
where $s^{[k]},a^{[k]},s^{[k+1]}$ are samples collected when the agent interacts with the environment in the $k$-th iteration and $\alpha^{[k]}$ is the step size. As the dimension of the Q table grows linearly in the cardinality of the state and action space, function approximation is introduced to parameterize it in large-scale problems. In this work, we use a neural network to parameterize the Q function. We will use $\psi$ to denote the parameters of the neural network and $Q_{\psi}:\mathbb{R}^{|\Scal|\times|\Acal|}\rightarrow\mathbb{R}$ to denote the Q function parameterized by $\psi$. We also employ standard techniques such as double Q-learning \cite{van2016deep} and prioritized experience replay~\cite{schaul2015prioritized} to stabilize and accelerate training. More detailed introduction and theoretical treatment of Q learning can be found in \cite{watkins1992q,chen2019finite,zeng2021finite}.

\section{Algorithm Design}\label{sec:method}

In this section, we use the RL framework in Section \ref{sec:background} to develop a method that learns the penalty parameter $\rho$ in the ACOPF ADMM algorithm to accelerate its convergence.

While our objective is to reduce the number of ADMM iterations until convergence, the goal of an RL agent is to maximize the discounted cumulative reward it collects from the environment. To translate our objective to that of the RL agent, we have to model our ADMM parameter selection problem as a suitable RL problem, which includes identifying the environment and dynamics and making the proper choice of the state space, action space, and reward function.
\begin{figure}[h]
  \centering
  \includegraphics[width=.95\linewidth]{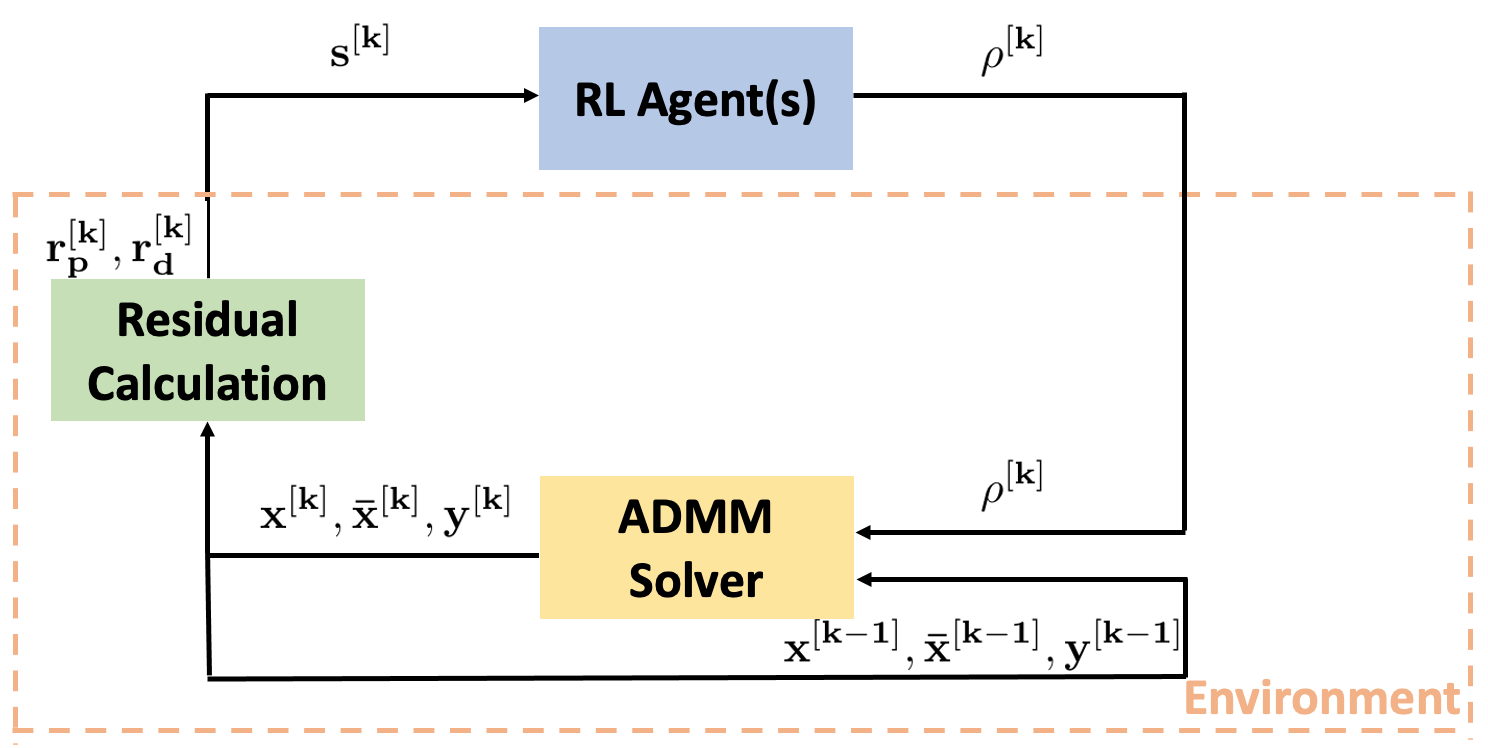}
  \caption{ADMM Solver and RL Agent Interaction}
  \label{figure:admm_rl}
\end{figure}

\subsection{RL Environment \& Reward Function}

We regard the ADMM solution process as the RL environment. Each iteration of the ADMM algorithm corresponds to one RL iteration. In iteration $k=0,1,\ldots$, the agent observes the current state of the ADMM solver $s^{[k]}$. Based on $s^{[k]}$, the agent selects an action $a^{[k]}$, which is simply a choice of $\rho^{[k]}$, the penalty parameter of the $k$-th iteration, and receives a reward $R(s^{[k]},a^{[k]})$, which we will design to reflect the value of the current state to the ADMM convergence.
The parameter $\rho^{[k]}$ is then fed back to the ADMM solver for another ADMM iteration. This process is repeated until both the primal and dual residuals from the ADMM solve drop below the thresholds, i.e., \eqref{eq:convergence}.
The interaction of the environment and the agent in ADMM solving process is shown in Figure~\ref{figure:admm_rl}.

\noindent\textbf{State space:}
The state is an important source of information that should summarize the progress of the ADMM algorithm and include key factors necessary for the agent to make decisions about $\rho$. 
In this problem, we naturally expect the primal and dual residuals to contain information about the optimal choice of $\rho$. To ensure that $s^{[k]}$ sufficiently represents the state of the ADMM solving process, we include the past $n$-point history of the residuals in $s^{[k]}$. 
\begin{align*}
    s^{[k]}\hspace{-2pt}=\hspace{-2pt}[(r^{[k-n+1]}_p,r^{[k-n+1]}_d),\cdots,(r^{[k]}_p,r^{[k]}_d)]\hspace{-2pt}\in\mathbb{R}^{2n\times(n_{pq}+n_{v\theta})}.
\end{align*}
\noindent\textbf{Action space:}
The algorithm used in this paper is a variant of Q-learning, which by design requires a discrete and finite action space. Since $\rho$ is only restricted to being positive, $\rho$ can be chosen from a continuous and infinitely large range of values. However, in the context of ACOPF problems, existing literature shows that $\rho$ values picked from a much smaller range result in superior convergence speed. Specifically, \cite{Mhanna19} suggests using two different $\rho$ for the two types of constraints: for constraints related to real and reactive power, $\rho_{pq}=400$ is used for IEEE 9-bus, 30-bus, and 118-bus systems; for constraints related to voltage, $\rho_{v\theta}=40000$ is used for IEEE 9-bus and 30-bus systems and $\rho_{v\theta}=4000$ is used for the 118-bus system. Though this particular choice of the parameters may not be optimal, it suggests a reasonable range for $\rho$ to provide to the RL agent.
We select $[100,1000]$ as the range of $\rho_{pq}$, and $[500,70000]$ for $\rho_{v\theta}$ in the 9-bus and 30-bus systems and $[500,7000]$ in the 118-bus system, discretized to 10 possible actions for each constraint (see Table~\ref{table:action_space}). We note that the action space may be more challenging to design for power networks in which we lack extensive prior knowledge. One solution for such networks is to determine by trial and error reasonable $\rho_{pq}$ and $\rho_{v\theta}$ values that lead to convergence, as commonly practiced in the existing works in distributed OPF \cite{mhanna2018component,Mhanna19}. Then an action space can be formed around these values.
\begin{table}[!h]
\centering
\caption{RL Action Space ($\rho$) \& Initial $\rho$ Values}
\begin{tabular}{ccp{4.2cm}}
        \toprule
        $\rho$ Category & Initial Value & \hspace{35pt}Action Space\\
        \midrule
        $\rho_{pq}$  & 400 & 
        \{100, 200, 300, 400, 500, 600, 700, 800, 900, 1000\}\\
        \midrule
        $\rho_{v\theta}$ (9-, 30-bus) & 40000 & \{500,~2000, ~5000,~10000,~20000, 30000,~40000,~50000,~60000,~70000\}   \\
        \midrule
        $\rho_{v\theta}$ (118-bus) & 4000 &  \{500, 1000, 1500, 2000, 2500, 3000, 3500, 4000, 5500, 7000\}\\
        \bottomrule
        \bottomrule
        \end{tabular}
\label{table:action_space}
\end{table}

\noindent\textbf{Reward function:}
The reward function is a crucial signal that affects the behavior of the agent. We have to carefully design the reward function to translate our objective, which is to accelerate ADMM convergence, correctly to the agent. The reward function $R$ should be chosen such that $R(s,a)$ is large if taking action $a$ while in state $s$ leads to fast convergence and small if taking action $a$ while in state $s$ leads to slow convergence.
With this in mind, a natural choice of the reward function is a large bonus given only to the convergence state; for instance,
\begin{align*}
    R_{\text{conv}}(s^{[k]},a^{[k]})=\begin{cases} 200, & \text{\hspace{-6pt}if }\hspace{-2pt}\left\|r^{[k+1]}_p\right\|_{2}\hspace{-2pt}\leq\hspace{-2pt}\epsilon_{p}\hspace{-2pt}\text{ and }\hspace{-3pt}\left\|r^{[k+1]}_d\right\|_{2}\hspace{-2pt}\leq\hspace{-2pt}\epsilon_{d}, \\ 0, & \text {\hspace{-6pt}else. }\end{cases}
\end{align*}
Due to the presence of the discount factor $\gamma\in(0,1)$, the reward received further in the future becomes less valuable. Therefore, to maximize the discounted cumulative reward under this reward function, the agent will aim to reach the convergence state in as few iterations as possible.

\begin{algorithm}[t]
\caption{Parameter Learning Through Q-Learning in ADMM ACOPF Solver}
\label{Alg:rl_admm}
\begin{algorithmic}[1]
\STATE{\textbf{ADMM initialization:} Initial parameters $x^{[0]}\in\mathbb{R}^{n_1}, \bar{x}^{[0]}\in\mathbb{R}^{n_1},y^{[0]}\in\mathbb{R}^{n_3}, \rho^{[0]}\in\mathbb{R}^{n_3}$}
\STATE{\textbf{RL initialization:} Initial Q function parameter $\psi^{[0]}$, \\
step size sequence $\alpha^{[k]}$, greedy policy parameter sequence $\epsilon^{[k]}$, length of state vector $n$, action space $\Acal$}
\FOR{$k=0,1,2,...$}
\IF{$k\geq n$}
    \STATE{Compute residuals $r^{[k]}_d$, $r^{[k]}_p$ from $x^{[k]},\bar{x}^{[k]}$ and\\
    form state vector $s^{[k]}=[(r^{[k-n+1]}_p,r^{[k-n+1]}_d),\cdots,(r^{[k]}_p,r^{[k]}_d)]$}
    \STATE{Select action $a^{[k]}\hspace{-2pt}\sim\hspace{-2pt}\widehat{\pi}^{[k]}(\hspace{-1pt}\cdot\hspace{-2pt}\mid \hspace{-2pt}s^{[k]})$ and translate to $\rho^{[k]}$}
\ELSE
    \STATE{Use the initial $\rho$ value: $\rho^{[k]}=\rho^{[0]}$}
\ENDIF
\STATE{Perform an ADMM update \eqref{eq:admm_update} with the current penalty parameter $\rho^{[k]}$}
\IF{$k\geq n$}
    \STATE{Receive reward $R(s^{[k]},a^{[k]})$, observe the next \\
    state $s^{[k+1]}$, and compute the Q target
    \begin{align*}
        Q_{\text{target}}=R(s^{[k]},a^{[k]})+\max_{a}Q_{\psi^{[k]}}(s^{[k+1]},a)
    \end{align*}}
    \STATE{Update the Q function parameter
    \begin{align*}
        \hspace{-8pt}\psi^{[k+1]}\hspace{-2pt}=\hspace{-2pt} \psi^{[k]}\hspace{-2pt}-\hspace{-2pt}\alpha^{[k]}\nabla_{\hspace{-2pt}\psi}(Q_{\psi}(s^{[k]},a^{[k]})\hspace{-2pt}-\hspace{-2pt}Q_{\text{target}})^2 \hspace{-2pt}\mid_{\psi=\psi^{[k]}}
    \end{align*}}
    \STATE{Update the $\epsilon$-greedy policy
    \begin{align*}
        \widehat{\pi}^{[k+1]}(a\mid s)\hspace{-2pt}=\hspace{-2pt}\begin{cases} 1\hspace{-2pt}-\hspace{-2pt}\frac{(|\Acal|-1)\epsilon^{[k]}}{|\Acal|}, & \hspace{-10pt}\text { if } a\hspace{-2pt}= \hat{a}^{[k+1]}(s)\\ \frac{\epsilon^{[k]}}{|\Acal|}, & \hspace{-10pt}\text { otherwise }\end{cases}
    \end{align*}
    where $\hat{a}^{[k+1]}(s)=\argmax_{a}Q_{\psi^{[k+1]}}(s,a)$.
    }
\ENDIF
\STATE{\textbf{Terminate} if ADMM has converged}
\ENDFOR
\end{algorithmic}
\end{algorithm}

Though this design of the reward function encodes our objective very well, it causes the agent to receive extremely sparse reward signals in the training process. Until the very last iteration, the agent will not receive any useful reward throughout the hundreds or thousands of iterations that are typically required for ADMM algorithms to converge for moderately sized ACOPF problems. Sparse rewards commonly cause exploration and credit assignment issues in RL \cite{hare2019dealing} and significantly slow down the learning process.


To offer a denser signal to the RL agent, we add the residuals in the reward function. Specifically, the reward received by the agent in state $s^{[k]}$ is proportional to the reduction in $\|r^{[k+1]}_p\|_2$ and $\|r^{[k+1]}_d\|_2$ from $\|r^{[k]}_p\|_2$ and $\|r^{[k]}_d\|_2$:
\begin{align*}
    &R_{\text{res}}(s^{[k]},a^{[k]})\notag\\
    &\hspace{10pt}=\frac{1}{Z_p}(\|r^{[k+1]}_p\|_2-\|r^{[k]}_p\|_2)+\frac{1}{Z_d}(\|r^{[k+1]}_d\|_2-\|r^{[k]}_d\|_2),
\end{align*}
where $Z_p$ and $Z_d$ are normalizing factors that balance the magnitude difference between the primal and dual residuals. This reward function makes sense, as achieving fast convergence is equivalent to quickly driving the residuals to the thresholds. This reward is non-zero in every ADMM iteration.

While we observe that the combination of $R_{\text{conv}}$ and $R_{\text{res}}$ works well in this problem, we further innovate the reward function design by taking advantage of the non-counterfactual nature of the environment.
We note that in most RL problems, the environment transition is irreversible, that is, once an action $a^{[k]}$ is deployed in state $s^{[k]}$, the environment moves forward to the next state $s^{[k+1]}$, and the consequence of selecting a different action in $s^{[k]}$ is never observable. However, in this problem, the progress of every ADMM iteration can be saved and we can therefore try different actions in the same state and compare their outcomes.
This feature of the environment affords more flexibility in the reward design.

In this work, we use a reward function computed with the help of a baseline policy $\tilde{\pi}$. In state $s^{[k]}$, we select the baseline action $\tilde{a}^{[k]}\sim\tilde{\pi}(\cdot\mid s^{[k]})$ and observe the resulting next state $\tilde{s}^{[k+1]}$ including primal and dual residuals $\tilde{r}^{[k+1]}_p$ and $\tilde{r}^{[k+1]}_d$. 
We note that this baseline action is only used to compute the residuals. We roll back to state $s^{[k]}$ once the residuals are collected. From state $s^{[k]}$, we then deploy the RL policy, making the environment transition to $s^{[k+1]}$ and reveal $r^{[k+1]}_p$ and $r^{[k+1]}_d$. The reward is defined as the relative advantage of the RL policy over the baseline:
\begin{align*}
    R_{b}(s^{[k]},a^{[k]})\hspace{-2pt}=\hspace{-2pt}\frac{\|r^{[k+1]}_p\|_2\hspace{-2pt}-\hspace{-2pt}\|\tilde{r}^{[k+1]}_p\|_2}{\|\tilde{r}^{[k+1]}_p\|_2}\hspace{-2pt}+\hspace{-2pt}\frac{\|r^{[k+1]}_d\|_2\hspace{-2pt}-\hspace{-2pt}\|\tilde{r}^{[k+1]}_d\|_2}{\|\tilde{r}^{[k+1]}_d\|_2}.
\end{align*}
This reward function essentially aims to achieve the same goal as $R_{\text{res}}$, but can have much smaller variance. To see this, note that $\|r^{[k+1]}_p\|_2-\|r^{[k]}_p\|_2$ and $\|r^{[k+1]}_d\|_2-\|r^{[k]}_d\|_2$ can fluctuate across several orders of magnitude  through ADMM iterations regardless of the choice of $\rho$. The reward function $R_{b}$ effectively removes the impact of the natural fluctuation of the residuals and makes the variance of $R_b$ significantly smaller than that of $R_{\text{res}}$. It has been observed that reducing the variance of the reward is critical in accelerating learning and is also the motivation behind popular algorithms such as the advantage actor-critic (A2C) \cite{mnih2016asynchronous}.
We emphasize that the sole purpose of the baseline policy is to offset the fluctuation in the norm of the residuals over iterations. Therefore, the baseline policy can be very simple. In the experiments of this work, the baseline policy is to always use $\rho_{pq}=500$ and $\rho_{v\theta}=500$. Accordingly, the reward function we choose in this work combines $r_{\text{conv}}$ and~$r_b$:
\begin{align*}
    R(s^{[k]},a^{[k]})=R_{\text{conv}}(s^{[k]},a^{[k]})+R_{b}(s^{[k]},a^{[k]}).
\end{align*}


\subsection{Factorized Entry-wise Policy}\label{sec:factorizedpolicy}

We have discussed the transformation of the ADMM parameter selection problem into a RL problem where the policy selects a vector $\rho$ given the state vector. With the ten possible choices of $\rho$ values for each constraint, the total cardinality of the action space is $10^{n_{pq}+n_{v\theta}}$, which grows exponentially in the number of constraints and quickly becomes computationally intractable. To address this issue, we reduce the action space by simplifying the policy using its structure. 

We observe that the dimension of the $\rho$ vector is equal to the number of constraints. 
Let $\pi_i$ be the policy for updating parameter $\rho_i$ with respect to the constraint $i$.
We assume that each policy function (i.e., conditional probability distribution) is independent of the others.
Then, if every entry of the state vector contains enough information to optimally determine the corresponding entry of $\rho$, the policy can be factorized as
\begin{align*}
    \pi(a\mid s)=\prod_{i=1}^{n_{pq}+n_{v\theta}}\pi_{i}(a_i\mid s_i),
\end{align*}
which means that we can equivalently train smaller policies $\pi_i$ for each $i=1,\dots,(n_{pq}+n_{v\theta})$, whose effective action space has a cardinality of 10. 
Learning the set of small policies with its size scaling up linearly in the number of constraints, however, can still be computationally expensive.
Therefore, we take one more step to simplify the policy by assuming that there exists two entry-wise policies $\pi_{pq}$ and $\pi_{v\theta}$ that can optimally determine the mappings from the entries of state vector to the entries of $\rho$ for all power and voltage constraints, respectively. This means that the policy can be further factorized as
\begin{align*}
    \pi(a\mid s)=\prod_{i=1}^{n_{pq}}\pi_{pq}(a_{pq,i}\mid s_{pq,i})\prod_{i=1}^{n_{v\theta}}\pi_{v\theta}(a_{v\theta,i}\mid s_{v\theta,i}).
\end{align*}
As a result of this factorization, we only need to learn and maintain two small entry-wise policies. Since we use the Q-learning algorithm in this work with an action space of size 10 for each constraint, this amounts to learning two Q functions $Q_{pq},Q_{v\theta}:\mathbb{R}^{2n}\times \mathbb{R}^{10}\rightarrow\mathbb{R}$.

In the ACOPF ADMM algorithm, we expect it to be generally impossible to determine the optimal $\rho$ entry for a particular constraint without information from the other constraints. Moreover, there may not exist two unified policies $\pi_{pq}$ and $\pi_{v\theta}$ that work optimally for all power and voltage constraints. However, simplifying the policy in this manner effectively reduces the learning complexity, and as we will show in Section~\ref{sec:experiments}, the policy pair $(\rho_{pq},\rho_{v\theta})$ achieves good empirical performance.

Along with advantages in computational tractability, another important benefit of the factorized entry-wise policy lies in its ability to be deployed to ACOPF ADMM problems with different numbers of constraints from the one seen by the RL agent in training. This means that the entry-wise policy pair trained under one power network can be flexibly applied to various other network structures. Later in Section \ref{sec:experiments}, we will discuss an important generalization of the learned policy to minor system modifications, where it is necessary for the policy to adapt to a change in the number of constraints.

\section{Numerical Experiments}\label{sec:experiments}


We demonstrate the performance of our RL model by training the parameter selection policy and testing its performance on the 9-bus, 30-bus, and 118-bus IEEE networks in the M{\sc atpower} format~\cite{matpower}.
Two additional evaluation tasks are carried out to validate the generalization of the learning performance to the practical scenarios in power system operations. In the first task, the RL policy is
evaluated for its effectiveness in unseen load profiles in the original network. This is an important task as the loads of a power system change frequently, requiring the ACOPF problem to be solved repeatedly in an efficient way.
The second task 
tests the RL policy on a slightly modified version of the system by removing generators and/or disconnecting transmission lines. This task is more challenging and also important in practice since we need to solve ACOPF problems under generator and line outages. 

Two small-sized neural networks of identical structure (4 fully-connected layers with hidden dimension 256) are used to approximate $Q_{pq}$ and $Q_{v\theta}$. The action space has dimension 10, and we choose the number of residual history points $n=20$. This makes the input and output dimension of the neural network 40 and 10, respectively. We take the initial $\rho_{pq}$ and $\rho_{va}$ to be the values  suggested by \cite{Mhanna19} (provided in Table \ref{table:action_space}). Each test instance is solved from a cold-start in ADMM.

\begin{table}[]
\centering
\caption{ADMM Iterations of RL Policy Under Training Loads}
\begin{tabular}{l|c|c|c|}
\cline{2-4}
& {[}Mhanna 2019{]} & RL policy & \begin{tabular}[c]{@{}c@{}}Iteration Reduction\end{tabular} \\ \hline
\multicolumn{1}{|l|}{9-bus} & 879 & 358 & 59.3\%  \\ \hline
\multicolumn{1}{|l|}{30-bus} & 1400 & 738 & 47.3\% \\ \hline
\multicolumn{1}{|l|}{118-bus} & 525 & 343  & 34.7\%  \\ \hline
\end{tabular}
\label{table:default_load}
\end{table}

\subsection{Performance on Training Scheme}\label{sec:exp_training}

The RL policy is trained under the default loading for 1000 RL episodes, where one episode is a complete ADMM solution process. Compared with the state-of-the-art $\rho$ adjustment scheme in \cite{Mhanna19} that results in ADMM convergence in 879, 1400, and 525 iterations for 9-bus, 30-bus, and 118-bus systems, the RL policy reduces the number of ADMM iterations by at least 30\% (Table \ref{table:default_load}). To understand the mechanism behind the fast convergence under the RL policy, we show the primal and dual residuals over ADMM iterations under the RL policy and the scheme in\cite{Mhanna19} for the 9-bus system. While the scheme in \cite{Mhanna19} leads to frequent fluctuations of the residuals which prolong the ADMM solving process, the RL policy avoids these fluctuations. Although this trend is not as obvious in 30-bus and 118-bus systems, we still observe that the RL policy allows the residuals to drop more smoothly.

\begin{figure}[h]
  \centering
  \includegraphics[width=.99\linewidth]{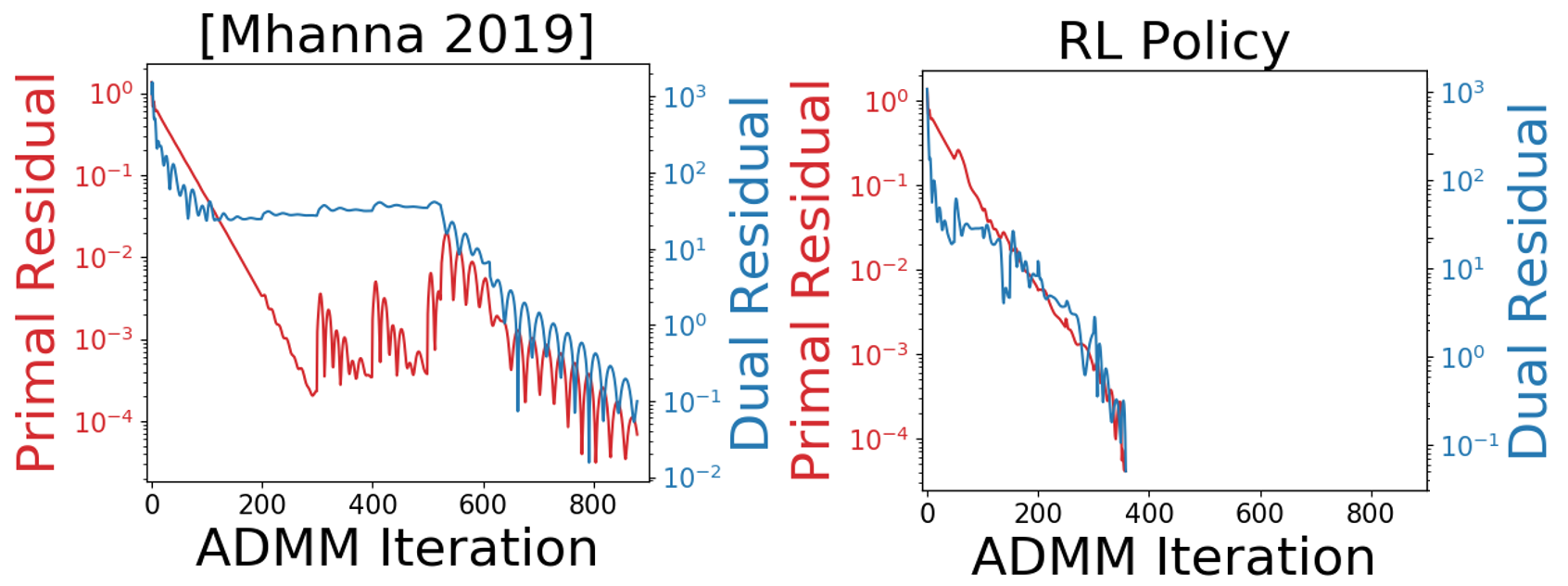}
  \caption{Convergence of Residuals with RL Policy for the 9-bus System}
  \label{figure:9bus_residuals}
\end{figure}



\subsection{Generalization of RL Policy to Varying Loads}\label{sec:exp_loadchange}

We also test the generalization of the RL policy to varying loads. 
Note that the RL policy has only been trained on the default loads from M{\sc atpower}, not on any other loading schemes.
We create a dataset of 50 test instances by randomly perturbing the default loads in the range $[-10\%, 10\%]$ at each bus. We summarize the number of ADMM iterations to convergence in Table~\ref{table:varyingloads}. The RL policy reduces the ADMM iterations relative to the scheme in \cite{Mhanna19} by 28\% to 50\% across test cases.

\begin{table}[!ht]
\centering
\caption{ADMM Iterations of RL Policy Under Varying Loads}
\begin{tabular}{l|c|c|c|c|c|}
\cline{2-6}
& \multicolumn{4}{c|}{$\rho$ selection method}  &  \\ \cline{2-5}
& \multicolumn{2}{c|}{{[}Mhanna 2019{]}} & \multicolumn{2}{c|}{RL policy}    &    \\ \cline{2-5}
& mean   & std   & mean & std & \multirow{-3}{*}{\begin{tabular}[c]{@{}c@{}}Iteration Reduction\end{tabular}} \\ \hline
\multicolumn{1}{|l|}{9-bus}  & 813.4 & 20.4 & 407 & 9.9 & 50.0\%\\ \hline
\multicolumn{1}{|l|}{30-bus}  & 1414.3 & 43.6 & 772.5 & 18.9 & 45.4\% \\ \hline
\multicolumn{1}{|l|}{118-bus} & 486.6 & 8& 346 & 7.2  & 28.9\%  \\ \hline
\end{tabular}
\label{table:varyingloads}
\end{table}

\subsection{Generalization of RL Policy to Generator and Line Outages}\label{sec:exp_outage}

In practical situations, we may need to solve the ACOPF problem after generator and line outages. Thus, it is of interest to investigate the performance of the RL policy in a modified network. 
In this section, we evaluate the ADMM convergence speed when applied to systems with 1) one generator removed and 2) one line disconnected.\footnote{
We consider all possible generator outage scenarios. Line outages are sampled in a uniformly random manner such that they do not island the network. We exclude line outages that lead to infeasible solutions under the method in~\cite{Mhanna19}. 
}
Again, we note that the RL policies were trained on the original M{\sc atpower} networks, without considering line or generator losses.
Tables~\ref{table:generator_outage} and~\ref{table:line_cutoff} summarize the performance of the RL policy and its comparison with the state-of-the-art method in~\cite{Mhanna19}.

\begin{table}[!ht]
\centering
\caption{ADMM Iterations of RL Policy Under Generator Outages}
\begin{tabular}{l|c|c|c|c|c|c|}
\cline{2-7}
&   &  \multicolumn{4}{c|}{$\rho$ selection method}  &  \\ \cline{3-6} & &  \multicolumn{2}{c|}{{[}Mhanna 2019{]}} & \multicolumn{2}{c|}{RL policy}  & \\ \cline{3-6}
& \multirow{-3}{*}{\begin{tabular}[c]{@{}c@{}}No. \\ of instances\end{tabular}} & mean    & std   & mean & std & \multirow{-3}{*}{\begin{tabular}[c]{@{}c@{}}Iteration \\ Reduction\end{tabular}} \\ \hline
\multicolumn{1}{|l|}{9-bus}  & 3 & 856.0 & 221.4 & 654.0  & 119.9 & 23.6\%    \\ \hline
\multicolumn{1}{|l|}{30-bus}  & 6 & 1325.8 & 404.3 & 695.8  & 78.9 & 47.5\% \\ \hline
\multicolumn{1}{|l|}{118-bus} & 54  & 483.8   & 17.7  & 340.0     & 8.8  & 29.7\% \\ \hline
\end{tabular}
\label{table:generator_outage}
\end{table}

\begin{table}[!ht]
\centering
\caption{ADMM Iterations of RL Policy Under Line Outages}
\begin{tabular}{l|c|c|c|c|c|c|}
\cline{2-7}
&   &  \multicolumn{4}{c|}{$\rho$ selection method}  &  \\ \cline{3-6} & &  \multicolumn{2}{c|}{{[}Mhanna 2019{]}} & \multicolumn{2}{c|}{RL policy}  & \\ \cline{3-6}
& \multirow{-3}{*}{\begin{tabular}[c]{@{}c@{}}No. \\ of instances\end{tabular}} & mean    & std   & mean & std & \multirow{-3}{*}{\begin{tabular}[c]{@{}c@{}}Iteration \\ Reduction\end{tabular}} \\ \hline
\multicolumn{1}{|l|}{9-bus}  & 6 & 698.7 & 218.5 & 367.3  & 31.1 & 47.4\%    \\ \hline
\multicolumn{1}{|l|}{30-bus}  & 10 & 1455.5 & 225.6 & 800.4  & 93.2 & 45.0\% \\ \hline
\multicolumn{1}{|l|}{118-bus} & 50  & 486.5   & 6.0  & 346.1     & 6.1  & 28.9\% \\ \hline
\end{tabular}
\label{table:line_cutoff}
\end{table}

\begin{figure}[h]
  \centering
  \includegraphics[width=.9\linewidth]{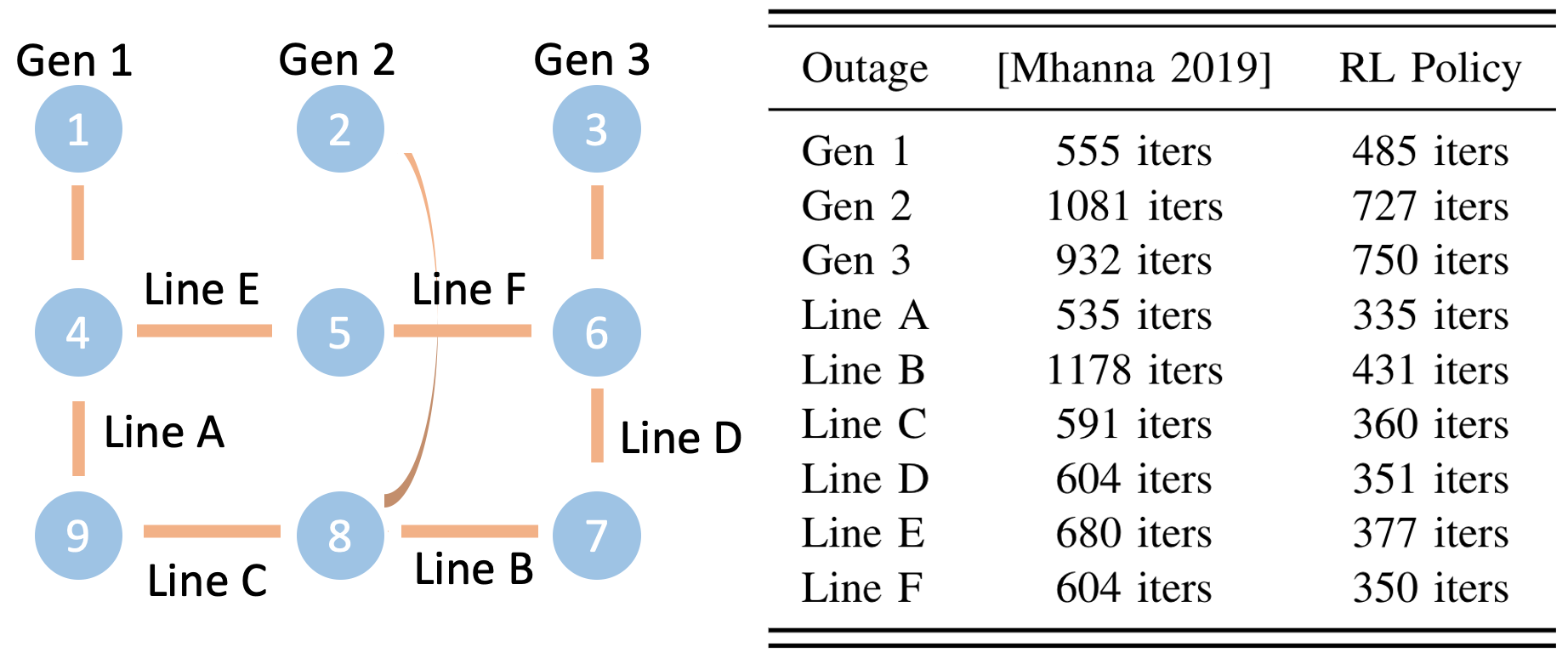}
  \caption{ADMM Convergence with RL Policy for the 9-bus System with Generator and Line Outages}
  \label{figure:9bus_networkcontingency}
\end{figure}

In the 9-bus system, there are three generator buses and six lines that can be disconnected while avoiding islands. In Figure~\ref{figure:9bus_networkcontingency}, we detail the ADMM convergence under the RL policy for each outage scenario, and note that the proposed method always outperforms \cite{Mhanna19} by a large margin.

\begin{remark}
The RL policy and the method in \cite{Mhanna19} achieve similar performance in terms of the converged objective function value in the experiments in Section~\ref{sec:exp_training} through Section~\ref{sec:exp_outage}. In most cases, the difference is within 1\%.
\end{remark}

\subsection{Generalization of RL Policy to Unseen Network Structures}

We also performed experiments on the generalization of the RL policy to networks that were not seen during training. For example, one may be interested in training a RL policy for a 9-bus system and deploying it to a 30-bus system. Accordingly, we trained RL policies for several systems and tested them on several others. Though our policy factorization described in Section~\ref{sec:factorizedpolicy} makes it possible to apply the RL policy to an ACOPF problem with a different number of constraints, experimentally, we found that policies trained in one network perform poorly in a completely different network. This observation strengthens our belief that there may not exist a universally optimal strategy that works for any ADMM problem, and thus supports the need for specialized approaches like the RL policies in this paper.

\begin{remark}
We note that the computational complexity of the RL policy for each constraint does not change with the dimension of system, as the dimension of the neural network is the same. However, the ADMM solver usually slows down as the system grows. Therefore, the computational time of deploying the RL policy should become increasingly negligible as the problem scales up. The training of the RL policy requires running 1000 episodes of complete ADMM solves, which involves a substantial amount of time but consists of offline computations that will not affect the online computational speed of deploying the RL policy.
In addition, while local primal and dual residuals need to be shared between computational nodes to train the RL policy, the policy runs completely locally once deployed, which reduces the leakage of local information during online operations. Mitigating privacy concerns in distributed OPF is an active research topic. We note that techniques in \cite{dvorkin2020differentially,ryu2021privacy} may help to further enhance the privacy preservation of our algorithm.
\end{remark}

\section{Conclusion \& Future Work}\label{sec:conclusion}

The choice of penalty parameters is key for accelerating the convergence of the ACOPF ADMM algorithm. By recognizing this task as a sequential decision making problem, we propose a RL framework in which we properly design the state space, action space, and reward function. We demonstrate the superior performance of the learned RL policy over the state-of-the-art method in a range of scenarios.
To the best of our knowledge, this is the first work to use machine learning for penalty parameter selection in distributed optimization for power systems applications. By reducing the number of ADMM iterations by up to 59\%,  this paper provides a successful proof of concept for using RL to enhance ADMM algorithms for power systems.

Future directions of the work include scaling up the method to larger systems and refining the RL training scheme to further improve ADMM convergence. 
We also plan to explore adapting this method to other distributed optimization algorithms beyond ADMM where local convergence can be theoretically guaranteed~\cite{lu2017fully,engelmann2018toward,sun2021two}.









\bibliographystyle{IEEEtran}
\bibliography{references}

\vspace{3mm}
{\footnotesize
\noindent\fbox{\parbox{0.47\textwidth}{
The submitted manuscript has been created by UChicago Argonne, LLC, Operator of Argonne National Laboratory (``Argonne''). Argonne, a U.S. Department of Energy Office of Science laboratory, is operated under Contract No. DE-AC02-06CH11357. The U.S. Government retains for itself, and others acting on its behalf, a paid-up nonexclusive, irrevocable worldwide license in said article to reproduce, prepare derivative works, distribute copies to the public, and perform publicly and display publicly, by or on behalf of the Government. The Department of Energy will provide public access to these results of federally sponsored research in accordance with the DOE Public Access Plan (http://energy.gov/downloads/doe-public-access-plan).}
}
}

\end{document}